\documentstyle[12pt]{article}
\begin{document}
\def\pb{\{.,.\}}
\def\om{\omega}
\def\omt{\tilde{\omega}}
\def\ti{\tilde}
\def\o{\Omega}
\def\Lm{\Lambda}
\def\bchi{\bar\chi^i}
\def\In{{\rm Int}}
\def\ba{\bar a}
\def\w{\wedge}
\def\ep{\epsilon}
\def\k{\kappa}
\def\Tr{{\rm Tr}}
\def\tr{{\rm tr}}
\def\ST{{\rm STr}}
\def\ss{\subset}
\def\ot{\otimes}
\def\bc{{\bf C}}
\def\br{{\bf R}}
\def\bt{{\bf T}}
\def\bz{{\bf Z}}
\def\bn{{\bf N}}
\def\de{\delta}
\def\mk{M_\k(\hat G)}
\def\mo{M_0(\hat G)}
\def\al{\alpha}
\def\la{\langle}
\def\ra{\rangle}
\def\G{{\cal G}}
\def\th{\theta}
\def\lm{\lambda}
\def\Th{\Theta}
\def\tt{\times_\Th}
\def\U{\Upsilon}
\def\e{\varepsilon}
\def\ve{\varepsilon}
\def\fe{{1\over \e}(1-{\rm e}^{-\e X^3})}
\def\ex{\js\e\bar XX}
\def\bq{\}_{qWZ}}
\def\thg{T^*\hat G}
\def\jp{{1\over 2}}
\def\tf{\hat t_{\infty}}
\def\tF{\hat T^{\infty}}
\def\ttf{\tilde t_{\infty}}
\def\ttF{\tilde T^{\infty}}
\def\zg{)_\G}
\def\hdg{\widehat{DG_0}}
\def\hlg{\widehat{LG_0}}
\def\lgc{LG_0^\bc}
\def\dgc{DG_0^\bc}
\def\ppp{{1\over 2\pi}}
\def\spp{{1\over 4\pi}}
\def\js{{1\over 4}}
\def\sp{{1\over 4\pi^2}}
\def\d{\partial}
\def\dz{\partial_-}
\def\dbz{\partial_+}
\def\LWD{Lu-Weinstein-Soibelman double}
\def\Dz{\partial_z}
\def\Dbz{\partial_{\bar z}}
\def\dto{\biggl({d\over ds}\biggr)_{s=0}}
\def\be{\begin{equation}}
\def\ee{\end{equation}}
\def\bea{\begin{eqnarray}}
\def\eea{\end{eqnarray}}
\def\D{{\cal D}}
\def\G{{\cal G}}
\def\H{{\cal H}}
\def\B{{\cal B}}
\def\E{{\cal E}}
\def\C{{\cal C}}
\def\A{{\cal A}}
\def\P{{\cal P}}
\def\R{{\cal R}}
\def\L{{\cal L}}
\def\wti{\widetilde}
\def\wh{\widehat}
\def\ttD{\tilde{\tilde D}}
\def\hhD{\hat{\hat D}}
\def\ttK{\tilde{\tilde K}}
\def\hhK{\hat{\hat K}}
\def\bbK{\bar{\bar K}}
\def\ttd{\tilde{\tilde \D}}
\def\hhd{\hat{\hat \D}}
\def\ttx{\tilde{\tilde x}}
\def\hhx{\hat{\hat x}}
\def\tty{\tilde{\tilde y}}
\def\hhy{\hat{\hat y}}
\def\tta{\tilde{\tilde \al}}
\def\hha{\hat{\hat \al}}
\def\ttb{\tilde{\tilde \b}}
\def\hhb{\hat{\hat \b}}
\def\ttxi{\tilde{\tilde \xi}}
\def\hhxi{\hat{\hat \xi}}
\def\ttyi{\tilde{\tilde \eta}}
\def\hhyi{\hat{\hat \eta}}
\def\hal{\hat\alpha}
\def\bl{\bar l}
\def\ptp{\stackrel{\otimes}{,}}
\def\hG{\hat {\cal G}}
\def\T{{\cal T}}
\def\F{{\cal F}}
\def\n{{1\over n}}
\def\dg{\dagger}
\def\si{\sigma}
\def\Si{\Sigma}
\def\RK{R^*_{P_K^{-1}}}
\def\l+{L_+G_0^\bc}
\def\+{D_+G_0^\bc}
\def\hdgc{\widehat{DG_0^\bc}}
\def\rn{\vert \alpha\vert^2}
\def\hlgc{\widehat{LG_0^\bc}}
\def\nl{{\nabla^L}}
\def\nr{{\nabla^R}}
\def\ot{\otimes}
\def\b{\beta}
\def\vr{\varrho}
\def\ga{\gamma}
\def\gc{G^{\bf C}}
\def\Gc{\G^{\bf C}}
\def\lw{D_{LWS}}
\def\LW{\D_{lws}}

\def\rdgc{\ ^\br\widehat {DG_0^\bc}}
\def\cdgc{\ ^\bc\widehat {DG_0^\bc}}
\def\rlgc{\ ^\br\widehat {LG_0^\bc}}
\def\clgc{\ ^\bc\widehat {LG_0^\bc}}

\def\st{\stackrel}
\def\stw{\stackrel{\w}{,}}
\def\0{^{01}}
\def\1{^{10}}
\def\od{\sqrt{2}}
\def\aq{[\al(\phi)]_\e}
\def\hbl{\hat\beta_L}
\def\hbr{\hat\beta_R}
\def\hmu{\hat\mu}
\def\bi{\bibitem}
\def\teo{\noindent {\bf Theorem}}
\def\defi{\noindent {\bf Definition}}
\def\rem{\noindent {\bf Remark}}
\def\pro{\noindent {\bf Proof}}
\def\lem{\noindent {\bf Lemma}}
\def\saf{{\e\over 2}\vert \hat a\vert^2\la i\hal^\vee,\hat \phi\ra}
\begin{titlepage}
\begin{flushright}
{}~
IML 01-xy\\
hep-th/0108148
\end{flushright}

\vspace{1cm}
\begin{center}
{\Huge \bf Quasitriangular chiral WZW model in a nutshell}\\
[50pt]{\small
{\bf Ctirad Klim\v{c}\'{\i}k}
\\ ~~\\Institute de math\'ematiques de Luminy,
 \\163, Avenue de Luminy, 13288 Marseille, France}
\end{center}

\vspace{0.5 cm}
\centerline{\bf Abstract}
\vspace{0.5 cm}
We give the bare-bone description of the quasitriangular
chiral WZW model for the particular choice of the
Lu-Weinstein-Soibelman  Drinfeld double   of the affine
Kac-Moody group. The symplectic structure of the model and
its Poisson-Lie symmetry
  are completely characterized by  two $r$-matrices
with spectral parameter. One of them is ordinary and
trigonometric and characterizes the $q$-current algebra. The other
is dynamical and elliptic (in fact  Felder's one) and characterizes
the braiding of  $q$-primary fields.
\end{titlepage}
\noindent {\bf 1. Introduction.} In reference \cite{K},
we have constructed a one-parameter deformation
of the standard WZW model. It is the theory possessing a
huge Poisson-Lie symmetry  generated by left and right $q$-deformed
current algebra.
The goal of this paper is to offer
the simplest  possible description of the $q$-deformed chiral
WZW theory  for the particular choice of the affine
Lu-Weinstein-Soibelman double.
 We address
those readers who want to obtain the first acquaintance with the
$q$-deformed WZW model by working with a representative simple example.
 Nothing will be derived
or proved here and only those results will be  presented
which do not require any preliminary understanding of the
Poisson-Lie world.
   In  particular,  we do not wish to bother the
reader with  more general choices of the Drinfeld double
 neither with  the natural origin of the
model from the symplectic reduction of a simpler system
living on the centrally extended double .
\vskip1pc
\noindent {2. \bf Language.}  The classical actions of dynamical
systems enjoying the
 Poisson-Lie
symmetry look typically forbiddingly complicated when written
in some of standard  parametrizations of
simple Lie groups.  This is the case already for toy
systems with few degrees of freedom, not even speaking about
field theories. It is therefore necessary to develop a language
suited for effective dealing  with such models.

 We can describe a classical
 dynamical system in more or less
 three different languages:
\vskip1pc
\noindent a) By defining a classical action on a space of fields.
This way is best suited for the path integral quantization.

\noindent b) By identifying a symplectic manifold and the Hamiltonian
function on it. This is good for the geometric quantization.

\noindent c) By picking up a representative set of (coordinate)
functions and Hamiltonian and defining their mutual Poisson brackets.
This is the starting point for the canonical quantization.

 \vskip1pc
We stress that at the classical level all three languages are fully
equivalent though in studying some particular feature of the system
one of them can turn out to be more convenient than the others.

The standard WZW model \cite{Wi} was invented in the language a).
The language b) was then extensively used for the description
of the (finite) Poisson-Lie symmetry of the chiral WZW model \cite{Gaw}.
Finally, the language c) has been also developed in papers devoted
to the canonical quantization of the (chiral) WZW model
\cite{AS,CGO,CG,CL}.

The quasitriangular WZW model \cite{K} was conceived by thinking
and working in  the language b).
However, the most  explicit (by  physicist's taste) and
economical description of this theory can be offered in the
 language c).
This is what we shall do here. The language a) in the
quasitriangular
case  in principle also exists. Indeed,
the interrelation between a) and b) is well-known. We recall
that for a couple (symplectic form $\omega$, Hamiltonian $H$)
we can immediately write the first order classical action of the
form
$$S=\int(d^{-1}\omega-Hdt).\eqno(1)$$
 The reader will agree, however, that the formula
(1) contains no additional insights with respect to b).  Moreover,
as we have already mentioned, the coordinate description of (1)
is awful.
We do not therefore expect usefulness of the path integral method
in the quasitriangular WZW story.
\vskip1pc
\noindent {\bf 3. Review of the standard chiral WZW model.}
Now we first review the definition of the
standard   WZW model \cite{CGO} in the language c).
Consider a compact, simple, connected and simply connected group
$G$.  Recall that the Weyl alcove $\A_+$
is the fundamental domain of the action of the Weyl group on the
maximal torus of $G$.
 Now the points of the phase space $P$
of the standard chiral WZW model  are the maps $m:\br\to G$,
fulfilling the
monodromy condition
$$ m(\si+2\pi)=m(\si)M.$$
Here the monodromy $M$ sits in the Weyl alcove $\A_+$ and it is
convenient to parametrize it as
 $$M=\exp{(-2\pi i a^\mu H^\mu)}.$$
 In other words,
$a^\mu$ are  coordinates on the alcove $\A_+$
corresponding to the choice of the orthonormal basis $H^\mu$ in
the Cartan subalgebra of $Lie(G)$.

The following matrix Poisson bracket (written in some
representation of $G$)
completely characterizes  the symplectic
structure   of the standard non-deformed chiral WZW model \cite{CGO}:
 $$ \{m(\si)\ptp m(\si') \}_{WZ} =(m(\si)\otimes m(\si'))
 B_{0}(a^\mu,\si-\si'), \eqno(2)$$
 where\footnote{Our conventions for the normalization of
the step generators $E^\al$ of $Lie(G)$ are as follows
 $$ [H^\mu,E^\al]=\la\al,H^\mu\ra E^\al,\quad(E^\al)^\dagger=
E^{-\al};$$
$$ \quad [E^\al,E^{-\al}]=\al^{\vee},\quad
[\al^\vee,E^{\pm\al}]=\pm 2E^{\pm\al},\quad
(E^\al,E^{-\al})_{\G_0^\bc}=
{2\over \vert \al\vert^2}.$$
The element $\al^\vee $ in the Cartan subalgebra of $Lie(G^\bc)$
is called the coroot of
 the root $\al$ and it is given by the formula
$$\al^\vee={2\over \vert \al\vert^2}\la \al,H^\mu\ra H^\mu.$$}
$$ B_0(a^\mu,\si)=
-{\pi\over\kappa}\biggl[\eta(\si)(H^\mu\ot H^\mu)-
i\sum_{\al}{\rn\over 2}{\exp{(i
\pi \eta(\si)\la \al,H^\mu\ra a^\mu)}\over
\sin{(\pi \la \al,H^\mu\ra a^\mu)}}
E^\al\ot E^{-\al}\biggr].$$
Here  $\k$ is the level and $\eta(\si)$ is the function defined by
$$ \eta(\si)=2[{\si\over2\pi}]+1,$$
with $[\si/2\pi]$ being the largest integer less than or equal
to ${\si\over 2\pi}$.

An important $Lie(G)$-valued observable is the chiral Kac-Moody current
$$j=\k mm^{-1}.\eqno(3)$$
It generates the hamiltonian action of (the central extension of)
the loop group $LG$ on the phase space $P$. This is reflected in the
following matrix Poisson brackets which  follow from (2) and (3):
$$\{m(\si)\ptp j(\si')\}_{WZ}=2\pi C \delta(\si-\si')(m(\si)\ot 1),
\eqno(4)$$
$$ \{j(\si)\ptp j(\si')\}_{WZ}=\pi\delta(\si-\si')[C, j(\si)\otimes
 1-1\otimes j(\si')]+\k  2\pi C\d_\si\delta(\si-\si'),\eqno(5)$$
where    $C$ is the Casimir element defined by
$$C=\sum_\mu H^\mu\ot H^\mu +\sum_{\al>0}{\vert\al\vert^2\over 2}
(E^{-\al}
\ot E^\al+E^{\al}\ot E^{-\al}).\eqno(6)$$
The Poisson bracket (5) corresponds to the non-deformed current algebra
and (4) can be interpreted as a statement, that $m(\si)$ is the
Kac-Moody
primary field. Finally, the relation (3) can be viewed as the
classical
version of the Knizhnik-Zamolodchikov equation \cite{KZ}.

The Hamiltonian of the non-deformed chiral theory is given by
the Sugawara formula:
$$H^{WZ}=-{1\over 2\k}(\k\d_\si mm^{-1},\k\d_\si mm^{-1})_{\G_0^\bc}.$$
It leads to the following simple time evolution in the phase space:
$$[m(\si)](\tau)=m(\si-\tau).$$
\vskip1pc
\noindent {\bf 4. q-current algebra}. The concept of the
$q$-deformation
of the current algebra (5) was apparently first introduced in
\cite{RST} who have worked out the complex case. The detailed
discussion
of the real case can be found in \cite{K}.  Here we shall need
only the
classical (Poisson bracket) story, which is based on the
 concept of a
meromorphic classical $r$-matrix $\hat r(\si)
\in Lie(G)\ot Lie(G)$ fulfilling
the ordinary classical Yang-Baxter equation
with  spectral parameter, i.e.
$$ [\hat r^{12}(\si_1-\si_2), \hat r^{13}(\si_1-\si_3)
+  \hat r^{23}(\si_2-\si_3)]+
[\hat r^{13}(\si_1-\si_3), \hat r^{23}(\si_2-\si_3)]=0.\eqno(7)$$
The $q$-current $L(\si)$ is a (hermitian in real case) matrix taking
values in some representation of $G^\bc$ and whose  Poisson
 brackets are
given by the formula
$$\{L(\si)\ptp L(\si')\}=(L(\si)\ot L(\si'))\e\hat r(\si-\si')
+\e\hat r(\si-\si')(L(\si)\ot L(\si')) $$
$$  -(1\ot L(\si'))\e\hat r(\si-\si'+2i\e\k)(L(\si)\ot 1)
-(L(\si)\ot 1)\e\hat r(\si-\si'-2i\e\k)(1\ot L(\si')).\eqno(8)$$
Here $\k$ is the level and $\e$ the deformation parameter related to
$q$ as $q=e^\e$.
Everywhere in this paper, $\hat r(\si)$ will denote the
following concrete
trigonometric solution of the YB equation (7):
$$\hat r(\si)=r+C{\rm cotg}\jp\si,$$
where $C$ is given by (6) and $r$ by
$$r=\sum_{\al>0}{i\vert\al\vert^2\over 2}(E^{-\al}\ot E^\al-E^\al\ot
E^{-\al}).$$
\vskip1pc
\noindent {\bf 5. Felder's elliptic dynamical $r$-matrix.}
 Consider the following $r$-matrix $B_\e(a^\mu,\si)
\in Lie(G)\ot Lie(G)$ (meromorphically) depending also  on the
 coordinates
$a^\mu$ of the alcove $\A_+$:
$$ B_{\e}(a^\mu,\si)=$$
$$ =-{i\over \k}\rho({i\si\over 2\k\e},{i\pi\over \k\e})
H^\mu\otimes H^\mu
-{i\over \k}\sum_{\al}{\rn\over 2}\si_{ a^\mu\la
\al,H^\mu\ra}
( {i\si\over 2\k\e},{i\pi\over \k\e})E^\al\otimes E^{-\al}.$$
 The elliptic functions $\rho(z,\tau),\si_w(z,\tau)$ are defined as
(cf. \cite{F,FW,EV})
$$ \si_w(z,\tau)={\theta_1(w-z,\tau)\theta_1'(0,\tau)\over
\theta_1(w,\tau)\theta_1(z,\tau)},
\quad \rho(z,\tau)={\theta_1'(z,\tau)\over \theta_1(z,\tau)}.
\eqno(9)$$
Note that  $\theta_1(z,\tau)$ is the Jacobi theta
function\footnote{We have
$\theta_1(z,\tau)=\vartheta_1(\pi z,\tau)$ with
$\vartheta_1$ in \cite{WHW}.}
$$ \theta_1(z,\tau)=-\sum_{j=-\infty}^{\infty}e^{\pi i (j+\jp)^2\tau
+2\pi i(j+\jp)(z+\jp)},$$
 the prime ' means the derivative with respect to the
 first argument $z$ and
the argument $\tau$ (the modular parameter )
is a nonzero complex number such that Im $\tau>0$.

It is straightforward to check (it helps to use \cite{WHW}), that
$B_\e(a^\mu,\si)$ verifies the following relations
$$ [B_{\e},1\otimes H^\mu +H^\mu\otimes 1]\equiv
[B_{\e}^{12},(H^\mu)^1 +(H^\mu)^2]=0;$$
$$ [B_{\e}^{12}(\si_1-\si_2), B_{\e}^{13}(\si_1-\si_3)
+  B_{\e}^{23}(\si_2-\si_3)]+
[B_{\e}^{13}(\si_1-\si_3), B_{\e}^{23}(\si_2-\si_3)]+$$
$$ +{i\over\k}( {\d\over\partial a^\mu}B_{\e}^{12})(H^\mu)^3+
{i\over\k}({\d\over\partial a^\mu}B_{\e}^{23})(H^\mu)^1+
{i\over\k}( {\d\over\partial a^\mu}B_{\e}^{31})(H^\mu)^2=0.\eqno(10)$$
Here (10) is called the dynamical Yang-Baxter equation
with spectral parameter \cite{F,EV}.

\vskip1pc
\noindent {\bf 6. The quasitriangular chiral WZW model.}
The phase space of the quasitriangular chiral WZW model is $P$;
i.e. it is the same as that of its non-deformed counterpart.
The symplectic structure of the deformed model is completely
characterized by the following matrix Poisson bracket:
$$  \{m(\si)\ptp m(\si') \bq =(m(\si)\otimes m(\si'))
 B_{\e}(a^\mu,\si-\si')+
 \e\hat r(\si-\si')(m(\si)\otimes m(\si')),\eqno(11)$$
where all notations have been already explained before.
The properties of the elliptic functions (9) imply that
$${\rm lim}_{\e\to 0} B_\e(a^\mu,\si)= B_0(a^\mu,\si).$$
This fact gives immediately the correct limit $q\to 1$
(cf. Eq. (2)).

The $q$-current $L(\si)$ is given by the   classical
version of the $q$-KZ equation:
$$L(\si)=m(\si+i\k\e)m^{-1}(\si-i\k\e).\eqno(12)$$
From (11) and (12), it    follows
$$ \{m(\si)\ptp L(\si')\}_{qWZ}=$$
$$ =\e \hat r(\si-\si'-i\e\k)(m(\si)\ot L(\si'))-
(1\ot L(\si'))\e \hat r(\si-\si'+i\e\k)
(m(\si)\ot 1)\eqno(13)$$
and
$$\{L(\si)\ptp L(\si')\}_{qWZ}=(L(\si)\ot L(\si'))\e\hat r(\si-\si')
+\e\hat r(\si-\si')(L(\si)\ot L(\si')) $$
$$  -(1\ot L(\si'))\e\hat r(\si-\si'+2i\e\k)(L(\si)\ot 1)
-(L(\si)\ot 1)\e\hat r(\si-\si'-2i\e\k)(1\ot L(\si')).\eqno(14)$$
The relation (13) can be interpreted that $m(\si)$ is the
$q$-primary field and (14) is nothing but the defining relation
(8) of the $q$-current algebra.

A few words about the $q\to 1$ limit:
From the classical $q$-KZ equation (12), we derive
$$ L(\si)= 1+2i\e\k \d_\si mm^{-1}+O(\e^2)=1+2i\e j(\si)+O(\e^2).$$
Inserting this into (13) and (14), we obtain in the lowest
order in $\e$ the desired relations (4) and (5):
$$ \{m(\si)\ptp j(\si')\}_{WZ}=2\pi C\delta(\si-\si')
(m(\si)\otimes 1);$$
$$ \{j(\si)\ptp j(\si')\}_{WZ}=\pi\delta(\si-\si')[C, j(\si)\otimes
 1-1\otimes j(\si')]+  2\pi\k C\d_\si\delta(\si-\si').$$
It turns out that the flow
$$[m(\si)](\tau)=m(\si-\tau)$$
on $P$ is Hamiltonian also for the $q$-deformed symplectic
structure (11).
Its
generator $H^{qWZ}$ is the Hamiltonian of the quasitriangular chiral
WZW model.  The explicit formula for it is given by the
 relation (1.20)
in \cite{K}. We do not list it here in order not to break the basic
promise expressed in the introduction: reading of
this  paper did not require any preliminary knowledge of the
 Poisson-Lie
world.
 
\end{document}